# Three-dimensional Helical-rotating Plasma Structures in Beam-generated Partially Magnetized Plasmas

Jian Chen[1, *], Andrew T. Powis[2], Igor D. Kaganovich[2], Zhibin Wang[1, *], Yi Yu[1, *]

[1]Sino-French Institute of Nuclear Engineering and Technology, Sun Yat-sen University, Zhuhai 519082, P. R. China

[2]Princeton Plasma Physics Laboratory, Princeton NJ 08540, USA

*Emails: chenjian5@mail.sysu.edu.cn , wangzhb8@mail.sysu.edu.cn and yuyi56@mail.sysu.edu.cn

Azimuthal structures emerging in beam-generated partially magnetized plasmas are investigated using three-dimensional particle-in-cell/Monte Carlo collision simulations. Two distinct instability regimes are identified at low pressures. When the gas pressure is sufficiently high, quasi-neutrality is attained and 2D spiral-arm structures form as a result of the development of a lower-hybrid instability, resulting in enhanced cross-field transport. At lower pressures, quasi-neutrality is not achieved and a 3D helical-rotating plasma structure forms due to development of the diocotron instability. Analytical formulas are proposed for the critical threshold pressure between these regimes and for the rotation frequency of the helical structures. Preliminary experimental verification is provided.

*Introduction* — Beam-generated partially-magnetized plasmas with magnetized electrons and unmagnetized ions have emerged as a promising technique for atomic-layer functionalization [1-3]. These plasmas often exhibit macroscopic azimuthally rotating structures with enhanced density, known as "spokes" [4-12]. The emergence of these structures results in significantly enhanced particle and energy transport across the magnetic field which can cause unexpected surface damage. Understanding the formation and dynamics of azimuthal structures is essential for precise control of beam-generated plasma properties.

Several instabilities have been identified as drivers of azimuthal structure formation. Krall et al. [13] and Hirose et al. [14] theoretically predicted a lower-hybrid instability induced by cross-field currents in nonuniform plasmas, which was later confirmed in beam-generated discharge experiments [15]. Another possible driver is the modified Simon-Hoh instability, originally studied by Simon [16] and Hoh [17]. For collisional cases, it is driven by the relative velocity drift between electrons and ions, arising in systems where the electric field and plasma density gradient satisfy the condition $\boldsymbol{E} \cdot \nabla n_0 > 0$. However, a modified Simon-Hoh instability exists in collisionless plasmas, where the difference in drift velocity arises due to the large Larmor radius of weakly magnetized ions [18]. Experimental observations of the modified Simon-Hoh instability have been reported in various beam-generated partially magnetized plasma devices [18-20]. More recently, the centrifugal instability, caused by inertial-induced differences in electron and ion azimuthal velocities, has also been proposed as a source of rotating structures [21, 22]. Additionally, a Turing-type activator-inhibitor model has been employed to interpret the azimuthally rotating filaments observed in the magnetized plasma device MDPX [23].

In addition to numerous experimental and theoretical efforts, numerical simulations have been indispensable in revealing the formation and dynamics of azimuthal structures during the nonlinear saturation stage [24-31]. While some preliminary 3D simulations have been presented recently [32, 33], most numerical studies focus on the 2D transverse dimension, neglecting longitudinal processes (parallel to the magnetic field). This omission overlooks important 3D effects, such as unstable longitudinal modes and the boundary effects observed in experiments [19, 34], underscoring the need for fully 3D kinetic simulations.

In this letter, we present 3D3V particle-in-cell/Monte Carlo collision (PIC/MCC) simulations of azimuthal structures in beam-generated partially magnetized plasmas. Beam electrons are injected into a grounded metal chamber along a uniform longitudinal magnetic field, ionizing neutral helium gas and producing a subsequent plasma. We identify two regimes, each controlled by different instabilities and their resulting azimuthal structures. These regimes are distinguished by whether the quasi-neutrality condition is attained. In the quasi-neutral regime, a lower-hybrid instability develops, forming quasi-2D spiral azimuthal arms. In the non-neutral regime, where ionization is insufficient to sustain quasi-neutrality, a diocotron instability arises, eventually leading to collective helical rotation of the entire plasma.

*Model* — The electron beam-generated plasma under a longitudinal magnetic field is modeled using the 3D3V LTP-PIC software, an extensively benchmarked [35, 36], explicit, electrostatic PIC/MCC code. The simulation domain consists of a three-dimensional Cartesian box with transverse dimensions $L_x=L_z=25$ mm and longitudinal length $L_y=200$ mm, with all boundaries grounded and fully absorbing. An electron beam is launched along the $+y$ direction from a small window (4 mm × 4 mm) at the center of the left boundary, with a density of $n_b=10^{15}$ m$^{-3}$, a temperature of $T_b=5$ eV and an energy of $\varepsilon_b=50$ eV. A uniform longitudinal magnetic field of $B=100$ G is applied and the domain is filled with helium gas at varying pressures. These parameters are motivated by the beam-generated plasma discharge experimental setup described in Refs. [34] and [37].

Electron-neutral collisions (elastic scattering, excitation, ionization) and ion-neutral charge exchange collisions are included in the simulations and modeled using the null-collision algorithm based on the Monte Carlo method [38, 39]. A cell size $\Delta x=390$ μm and a time step $\Delta t=20$ ps are specified to resolve the Debye length ($\lambda_{De}\approx525$ μm), plasma frequency ($\omega_{pe}\approx1.78\times10^9$ s$^{-1}$), and fulfill the Courant–Friedrichs–Lewy condition for beam electrons. The simulation is initialized with a plasma density of $n_0=1\times10^{15}$ m$^{-3}$ and 300 macro-particles per cell for each species. The simulations were run to steady state, (greater than 20 μs) using 4096 CPU cores on the Perlmutter supercomputer, typically completing within 2 days.

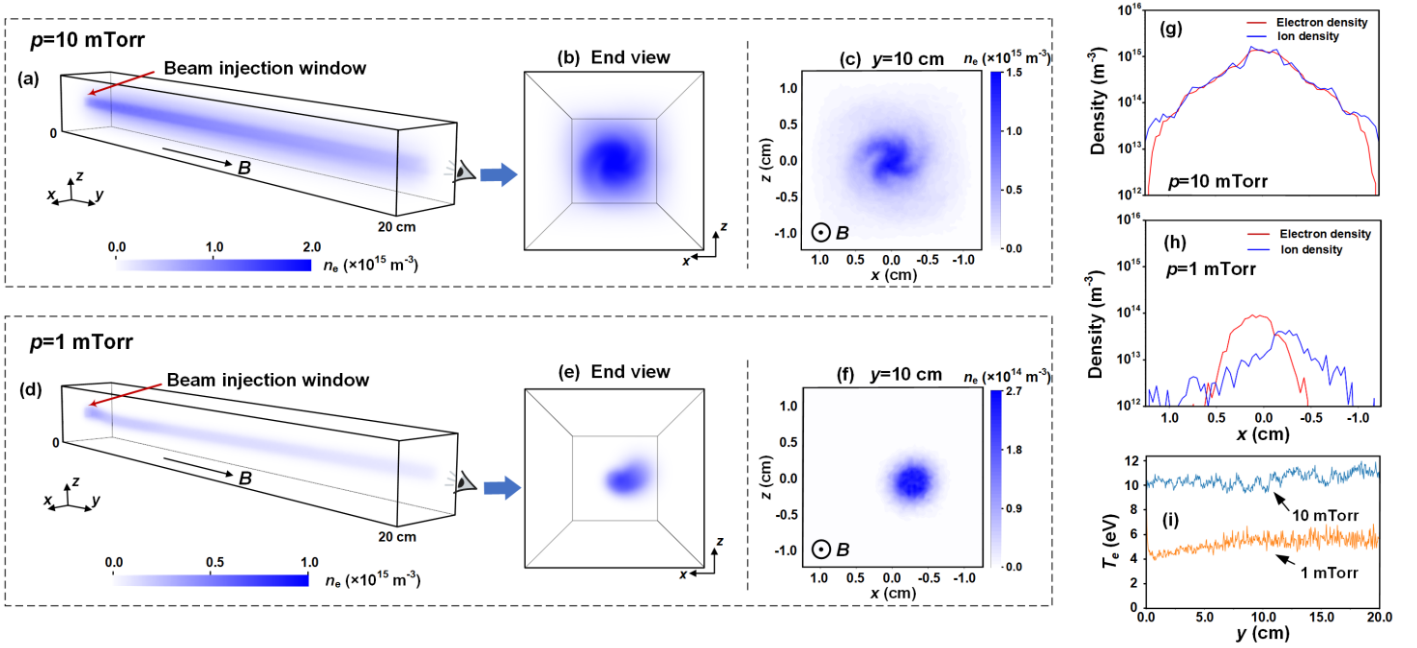

Fig. 1 Plasma parameters at quasi-steady state for $p$=10 mTorr and $p$=1 mTorr. Subfigures (a) and (d) show the 3D distribution of electron number density, with the magnitude illustrated in shaded blue. Subfigures (b) and (e) show the end views of (a) and (d) captured by a synthetic diagnostic "camera" near the end of the domain with a 30° view angle. Black and gray lines mark the 30° view angle boundaries. All 3D plots and end views are generated using the VisIt program [40]. Subfigures (c) and (f) show the transverse profiles of electron number density at $y$=10 cm. Subfigures (g) and (h) show the 1D profiles of electron and ion number densities along the line ($y$=10 cm, $z$=0 cm). Subfigure (i) shows the electron temperature profiles along the axis ($x$=0, $z$=0).

*Two regimes with distinct azimuthal structures*─ By varying the pressure, we identify two regimes characterized by distinct azimuthal structures, as depicted in Fig. 1. The plots correspond to a quasi-steady state, when particle production and losses are balanced and periodic azimuthal rotation persists.

As shown in Fig. 1(a), at $p$=10 mTorr, the rotating plasma primarily concentrates along the axis [($x$=0, $z$=0)]. In contrast, at $p$=1 mTorr, the plasma shifts off-axis, forming a three-dimensional helical-rotating structure [Fig. 1(d)] with an angular frequency $\omega \approx 5.6$ MHz and an eccentric distance of approximately 0.25 cm (see the Fourier spectrum in Section I of the Supplemental material and plasma dynamics in the Supplemental video [41]). This off-axis helical structure is suggestive of the electron-hose instability [42], or the transverse two-stream instability [43]. However, unlike the electron-hose and transverse two-stream instabilities, where transverse displacements oscillate in time and space, the instability observed here exhibits rotation as a 3D helical structure with nearly constant eccentricity.

To clarify the different azimuthal dynamics between the two cases, end-view images of the plasma [Figs.1(b) and 1(e)] and transverse electron density profiles [Figs.1(c) and 1(f)] are provided. In actual experiments, azimuthal dynamics are typically captured by fast cameras at the chamber's end window.

Therefore, we capture these images [Figs.1(b) and 1(e)] using a synthetic camera with a 30° view angle placed at the end of the domain, corresponding to the integration of electron density along the line of sight. At $p$=10 mTorr, Figs. 1(b) and 1(c) show azimuthal structures with curved tails extending from the center, a feature seen in previous radial-azimuthal simulations [26, 31]. At $p$=1 mTorr, the end-view image [Fig. 1(e)] also reveals spiral structures, resembling the $m$=1 azimuthal "arms" observed in linear plasma machines such as MISTRAL [8-11]. However, the transverse electron density profile in Fig. 1(f) indicates that these are not genuine 2D azimuthal structures but rather projections of an off-axis 3D helical structure.

These contrasting plasma behaviors suggest two distinct regimes. As shown in Figs. 1(g) and 1(h), the key distinction lies in whether quasi-neutrality is maintained. At $p$=10 mTorr, electron and ion density profiles are nearly identical (except in the sheath). However, at $p$=1 mTorr, quasi-neutrality breaks, with a clear displacement between electron and ion columns. This displacement along with electron beam rotation generates a time-dependent electric field that forces ions to collectively rotate, lagging behind the electrons. Both electron density and temperature are lower at $p$=1 mTorr [Figs. 1(g)~(i)], increasing the Debye length from $\lambda_{De}$≈0.6 mm to $\lambda_{De}$≈1.7 mm, however this remains significantly smaller than half of the transverse domain size. A detailed discussion on how non-neutrality is maintained at $p$=1 mTorr is provided in Section II of the Supplemental material [41]. In this Letter, we refer to the above two regimes as the quasi-neutral and non-neutral regimes.

*Dominant instabilities in the quasi-neutral and non-neutral regimes* — The azimuthal structures in the quasi-neutral and non-neutral regimes are generated as a result of different instabilities. For illustration, we analyze the first few microseconds of plasma evolution while instabilities develop and saturate.

In the quasi-neutral regime, as shown in Fig. 2, an $m$=4 azimuthal mode emerges with elongated and curved tails [Fig. 2(a)]. This mode arises from a lower-hybrid instability, destabilized by the radial density gradient, equilibrium $\boldsymbol{E}\times\boldsymbol{B}$ drift and collisions, shortly after beam injection. Given that $k_y$≈0, the dispersion relation for the lower-hybrid instability reads [31, 44]

$$k_\perp^2 \lambda_{De}^2 = \frac{k_\perp^2 c_s^2}{\omega^2} - \frac{\omega^* + k_\perp^2 \rho_e^2 (\omega - \omega_E + i\nu_{en})}{\omega - \omega_E + k_\perp^2 \rho_e^2 (\omega - \omega_E + i\nu_{en})}, \tag{1}$$

where $\omega^* = -mT_e/(eB_0 r L_n)$, $\omega_E = -mE_r/(rB_0)$, $L_n^{-1} = n_0'/n_0$ (the prime denotes the derivative in the radial direction), $k_\perp \approx k_\theta = m/r$ is the perpendicular wave number, $\omega_{ce} = eB_0/m_e = 1.76\times10^9$ s$^{-1}$ is the electron gyrofrequency, $n_0$ is the equilibrium number density, $E_r$ is the radial electric field, $\rho_{ce} = \sqrt{T_e/(m_e\omega_{ce}^2)}$ is the electron Larmor radius, $c_s = \sqrt{T_e/m_i}$ is the ion acoustic speed, and $\nu_{en}$ is the electron-neutral collision frequency. In the 2D limit of $k_y$≈0, beam and plasma electrons are assumed to have the same response to the transverse electric field perturbations. Fig. 2(b) presents the Fourier spectrum alongside the dispersion relation from Eq. (1). As shown, the observed $m=4$ mode matches the dispersion relation, suggesting that this mode is of the lower-hybrid type generated at the maximum growth rate. As the instability saturates, it stabilizes into the spiral azimuthal structures [Figs. 1(b) and 1(c)]. The local simulation parameters used in these calculations are summarized in Section III of the Supplemental material [41].

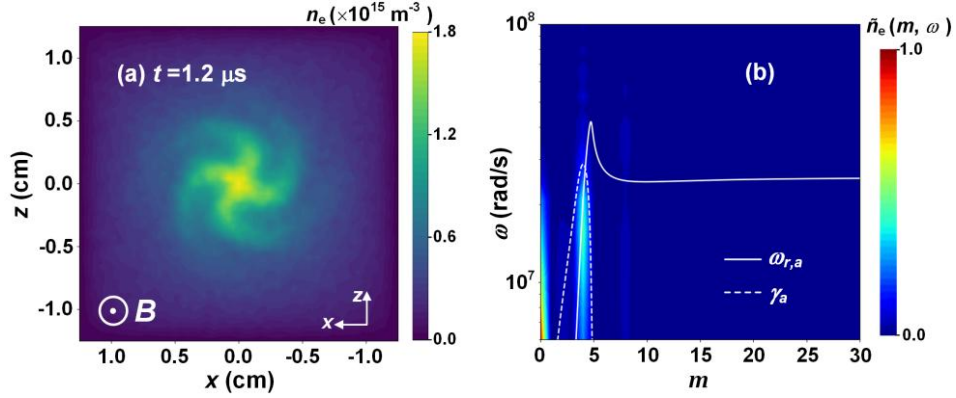

Fig. 2 (a) Electron number density profiles at $y=2$ cm and $t=1.2$ μs, showing the dominant $m=4$ azimuthal mode. The magnetic field vector is directed outward. (b) Fourier spectrum of azimuthal modes from 0 μs to 1.2 μs. The solid and dashed white lines denote the real frequency, $\omega_{r,a}$, and growth rate, $\gamma_a$, versus the azimuthal mode number $m$, calculated using Eq. (1). The data is taken from the case with $p=10$ mTorr.

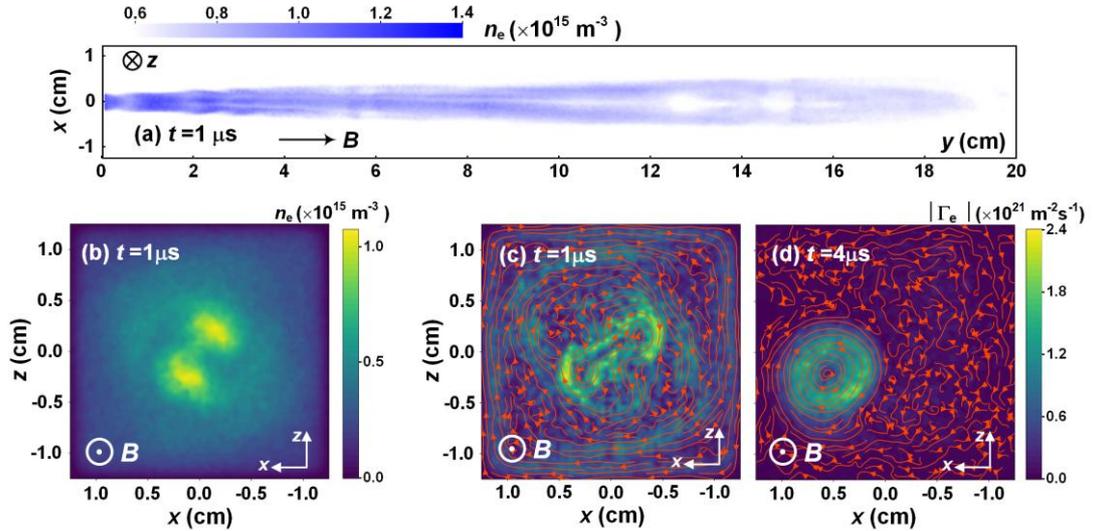

Fig. 3 (a) Transverse view of electron number density stereogram at $t=1$ μs; (b) Profile of electron number density at $y=10$ cm and $t=1$ μs; Streamlines of electron flux at $y=10$ cm at $t=1$ μs (c) and $t=4$ μs (d).

In contrast, plasma in the non-neutral regime evolves into two intertwined helical columns [Fig. 3(a)], each with a distinct density peak in the transverse profile [Fig. 3(b)]. Electrons near these peaks rotate around them, forming azimuthal vortices [Fig. 3(c)]. These two helical columns not only rotate collectively around the axis but also interact/compete with each other. Over time, one column fades, while the other grows larger and eventually evolves into an off-axis helical structure [see Fig. 3(d) and Fig. 1(d)]. The detailed evolution of this process can be found in Supplemental video [41].

Generation of the azimuthal vortices is the result of a *diocotron instability*- a Kelvin-Helmholtz-type instability arising from velocity shear generated by space charges [45], i.e., $\nabla \times \vec{v}_E = \frac{e(n_i - n_e)\vec{b}}{\varepsilon_0 B}$. Similar off-axis diocotron modes have been observed in Penning-Malmberg traps [46-51]. Our work presents, to our knowledge, the first 3D simulations showing the full evolution of a diocotron instability and its relationship to azimuthal structures.

The stability threshold for the diocotron instability can be calculated by analyzing the eigenvalue equation for the electrostatic potential perturbation, $\delta\phi(r)$, as derived in Refs. [45, 52-55]. Neglecting $v_{en}$, the eigenvalue equation for $\delta\phi(r)$ is given by

$$\frac{1}{r}\frac{\partial}{\partial r}\left[r\left(1-\frac{\omega_{pe}^2}{\eta_e^2}-\frac{\omega_{pi}^2}{\eta_i^2}\right)\frac{\partial}{\partial r}\delta\phi\right] - \frac{m^2}{r^2}\left(1-\frac{\omega_{pe}^2}{\eta_e^2}-\frac{\omega_{pi}^2}{\eta_i^2}\right)\delta\phi - k_y^2\left(1-\frac{\omega_{pe}^2}{(\omega-k_y v_y-\omega_E)^2}-\frac{\omega_{pi}^2}{\omega^2}\right)\delta\phi$$
$$= -\frac{m\delta\phi}{r(\omega-\omega_E)}\frac{\partial}{\partial r}\left[\frac{\omega_{pe}^2}{\eta_e^2}\left(-\omega_{ce}+2\frac{v_E}{r}\right)\right] - \frac{m\delta\phi}{r\omega}\frac{\partial}{\partial r}\left[\frac{\omega_{pi}^2}{\eta_i^2}\omega_{ci}\right], \quad (2)$$

where $\omega_{pe}$ and $\omega_{pi}$ are the electron and ion plasma frequencies, $v_y$ is the axial electron velocity, $v_E=r\omega_E/m$ is the azimuthal $E \times B$ drift velocity, and

$$\eta_e^2 = (\omega-k_y v_y-\omega_E)^2 - \left(-\omega_{ce}+2\frac{v_E}{r}\right)\left[-\omega_{ce}+\frac{1}{r}\frac{\partial}{\partial r}(rv_E)\right], \quad \eta_i^2 = \omega^2 - \omega_{ci}^2. \quad (3)$$

Although the diocotron mode has a finite $k_y$ (i.e., has a helical structure), $k_y \ll k_\theta$ and $k_y v_y \ll \omega_E$, so $k_y v_y$ can be neglected. In addition, considering that within the simulations $\omega_{ci}, \omega_{pi} \ll |\omega-\omega_E| \ll \omega_{ce}$, Eq. (3) is reduced to

$$\frac{1}{r}\frac{\partial}{\partial r}\left[r\frac{\partial}{\partial r}\delta\phi\right] - \frac{m^2}{r^2}\delta\phi = -\frac{m\delta\phi}{r(\omega-\omega_E)}\frac{e}{\varepsilon_0 B(1+\omega_{pe}^2/\omega_{ce}^2)}\left(\frac{\partial n_{e0}}{\partial r}-\frac{\omega_{pe}^2}{\omega_{ce}^2}\left(\frac{\partial n_{e0}}{\partial r}-\frac{\partial n_{i0}}{\partial r}\right)\right), \quad (4)$$

where $n_{e0}(r)$ is the equilibrium electron density. A similar equation was also obtained in Ref. [55]. As discussed in Refs. [52, 53, 55], for systems without a central conducting wall, instability occurs only if

$$\Lambda = \frac{\partial n_{e0}}{\partial r} - \frac{\omega_{pe}^2}{\omega_{ce}^2}\left(\frac{\partial n_{e0}}{\partial r}-\frac{\partial n_{i0}}{\partial r}\right) \quad (5)$$

changes sign as the radius increases from the center to the boundary. As shown in Section IV of the Supplemental material, for $p=1$ mTorr, $\Lambda$ crosses zero at $r \approx 0.32$ cm, indicating that the diocotron instability can be triggered at this radius (as confirmed by Fig. 3(b)).

We performed preliminary experimental verification of the diocotron mode in a newly developed linear plasma machine LEAD [56] (see Appendix A for a detailed discussion). We also develop a simple estimate for the rotation frequency of the helical structure based on its non-neutral nature (See Appendix B).

*Threshold pressure between the quasi-neutral and non-neutral regimes* — To estimate the threshold pressure delineating the quasi-neutral and non-neutral regimes, a global model accounting for ion balance is developed. Given that the observed radial ion density profile is exponential [Fig. 1(g)], the ionization balance yields

$$n_b v_{iz} S_{beam} \approx D_\perp n_{i0} \exp\left(-\frac{r}{|L_n|}\right) \cdot \frac{2\pi r}{|L_n|}, \tag{6}$$

where $v_{iz}$ represents the electron-impact ionization coefficient, $S_{beam}$ is the area of beam cross section, $n_{i0}$ is the ion number density at the axis, and $D_\perp$ is the perpendicular diffusion coefficient.

To solve Eq. (6), $D_\perp$ needs to be determined. As discussed in Section V of the Supplemental material [41], a large-amplitude fluctuating electric field near the axis significantly enhances radial diffusion. This instability-enhanced $D_\perp$ is challenging to estimate analytically since it is sensitive to both fluctuation levels and large-scale structure formation [57, 58]. Although Refs. [59, 60] provide analytical estimates of $D_\perp$ for radio-frequency magnetized discharges, the associated density profiles and ionization conditions differ significantly from our device, making their estimates not directly applicable to our system. To proceed, we consider recent experiments which provide a fitted $D_\perp$ using the form of a Bohm diffusion coefficient [37, 61]

$$D_\perp \approx \alpha \frac{T_e}{eB}. \tag{7}$$

where $\alpha$ is an anomalous parameter. Experimental observations show that $\alpha \approx 1/60$ near the axis and increases sharply with radius to maintain particle continuity. Although this fit was obtained at a pressure different from those in our simulations, $D_\perp$ is weakly dependent on pressure when a lower-hybrid instability determines the anomalous diffusion. Therefore, substituting Eq. (7) into Eq. (6) and setting $r=|L_n|$ yields

$$n_b v_{iz} S_{beam} \approx \frac{\pi}{30} \frac{T_e}{eB} n_{i0} \exp(-1). \tag{8}$$

Solving Eq. (8) yields $n_{i0}$. To satisfy the quasi-neutrality condition, $n_{i0}$ must exceed the beam density $n_b$. Consequently, the threshold pressure can be determined by equating $n_{i0}$ to $n_b$.

In addition to Eq. (7), $D_\perp$ can also be estimated via the modified Kadomtsev mixing-length transport coefficient [62-64]

$$D_\perp \approx \frac{2.4^2}{2} \left(\frac{\gamma_k}{k_\perp^2} \frac{\gamma_k^2}{\omega_k^2 + \gamma_k^2}\right)_{max}. \tag{9}$$

where $\gamma_k$ and $\omega_k$ are the linear growth rate and real frequency of the Fourier mode $k_\perp$. $k_\perp$ is typically chosen to maximize $D_\perp$. Solving Eq. (6) with Eq. (9) provides an alternative estimate of the threshold pressure.

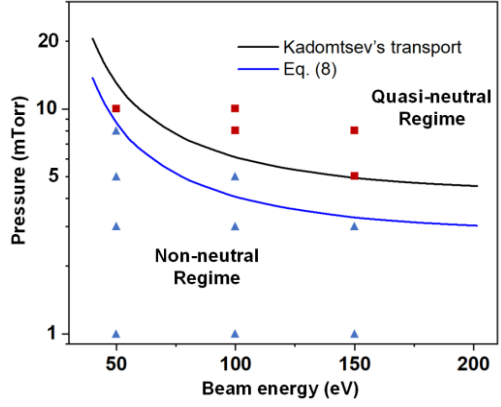

Fig. 4 Threshold pressure separating the quasi-neutral and non-neutral regimes versus beam energy. The blue line represents the solution of Eq. (8) with $T_e$=10 eV. The black line represents the solution by solving Eq. (6) with Eq. (9), using $k_\perp$=1600 m$^{-1}$, $\gamma_k$=2.9×10$^7$ s$^{-1}$ and $\omega_k$=1.6×10$^7$ determined from Fig. 2(b), along with the helium ionization cross section data from Ref. [65]. Red squares and blue triangles indicate simulation cases within the quasi-neutral and the non-neutral regimes, respectively.

To verify the predictions of threshold pressure, 12 additional 3D simulations were conducted across a range of pressures and beam energies, with the results shown in Fig. 4. The blue line represents the solution of Eq. (8), while the black line represents the solution obtained by solving Eq. (6) with Eq. (9). Kadomtsev's estimate has a nearly 50% difference compared to the experimental fit. The simulation results, denoted by red squares (quasi-neutral regime) and blue triangles (non-neutral regime), show good agreement with both the experimental fit and Kadomtsev's estimate. The deviation for some boundary cases suggests that the precise value of $D_\perp$ likely lies between the predictions of Eqs. (7) and (9).

*Conclusion*— This Letter presents 3D PIC/MCC simulations revealing two distinct regimes in beam-generated partially magnetized plasmas, each governed by different instabilities and their resulting azimuthal structures. In the quasi-neutral regime, the lower-hybrid instability generates spiral arms that govern the cross-field transport. In the non-neutral regime, the diocotron instability creates azimuthal vortices that disrupt the beam and result in collective helical plasma rotation.

We show that the axial projection of the helical-rotating plasma resembles spokes seen in experiments, which could lead to misinterpretation of this structure, thereby calling for further experimental investigation of 3D effects. We also propose and verify the analytical formulas for the threshold pressure delineating the two regimes and for the rotation frequency of the helical plasma, providing guidance for future experiments. Additionally, we present preliminary experimental evidence supporting the presence of azimuthal vortices.

*Acknowledgements*—The authors thank the referees for their careful review and helpful comments on the manuscript. The work of J. Chen and Z. B. Wang was supported by National Natural Science Foundation of China (Grant No. 12305223) and Guangdong Basic and Applied Basic Research Foundation (Grant No.

12305223). The work of A. T. Powis and I. D. Kaganovich at Princeton Plasma Physics Laboratory (PPPL) was supported by US Department of Energy under CRADA agreement between Applied Material Inc. and PPPL. Resources of the National Energy Research Scientific Computing Center (NERSC) were used for high performance computation. The work of Y. Yu was supported by the National Key Research and Development Program of China (No. 2022YFE03100002）.

# End matter

*Appendix A: Experimental verification of azimuthal vortex formation*—To validate the presence of the helical structure, we performed a preliminary experiment on the Linear Experimental Advanced Device (LEAD), a newly developed linear plasma machine at the Southwestern Institute of Physics in Chengdu, China. LEAD is primarily designed for the study of plasma turbulence physics and plasma-material interactions. A schematic of the LEAD device is presented in Fig. A1, with a detailed description provided in Ref. [56].

Currently, LEAD is equipped with a helicon plasma source, consisting of a four-ring copper antenna and a radio-frequency (RF) power supply. The RF power is introduced into the vacuum vessel through a quartz window at one end, generating a plasma column that is sustained and guided to the target chamber by an axial magnetic field. Although helicon plasmas exhibit differences from beam-generated plasmas, we argue that their behavior in $\mathbf{E}\times\mathbf{B}$ fields shares sufficient similarity to serve as a preliminary validation of the proposed effects.

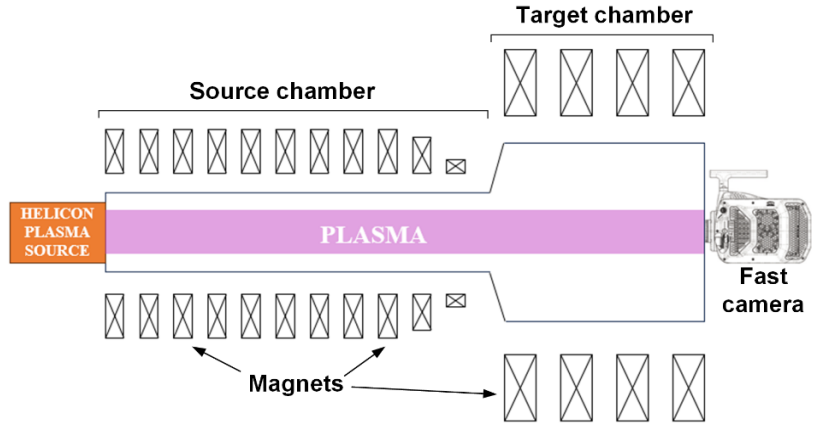

Fig. A1 Schematic of the LEAD device. The plasma is generated by a helicon source at one end of the source chamber. The source and target chambers have dimensions of ∅ 400 mm × 2 m and ∅ 900 mm × 1 m, respectively. A fast camera is placed at the end of the target chamber to capture cross-sections of the plasma column.

In the experiments, helium was used as a background gas with pressure of $p$=2 Pa. The RF power was set to $P$=4.5 kW and the axial magnetic field was maintained at $B$=900 G. These parameters differ from those employed in the simulations, however were selected to ensure the formation of a relatively uniform and stable plasma column in experiments. The primary objective of this study, at this initial stage, was to establish evidence for the presence of a helical structure. A more rigorous validation, aligning all conditions precisely with the simulations, will be pursued in future investigations.

To capture the azimuthal structures, a fast camera was placed at the end of the target chamber to observe a cross-section of the plasma column. The recorded snapshots, presented in Fig. A2, were captured with an exposure time of 7 μs and a shot-to-shot time interval of Δt=37 μs. As observed, the initial plasma column at t=0 rotates with respect to its centroid, remaining close to the axis. After around 3Δt, the plasma column begins to split and forms two distinct density peaks (see t=5Δt for two bright peaks). These peaks rotate in collective motion. Subsequently, one peak diminishes while the other remains, leading to the formation of an off-axis plasma column. This entire process closely resembles the observed azimuthal vortices and helical structures formed in our PIC simulations (see Fig. 3 in the main text). Therefore, these new experimental observations provide evidence supporting the presence of azimuthal vortices and helical structure in $E\times B$ plasmas, as seen in the simulations.

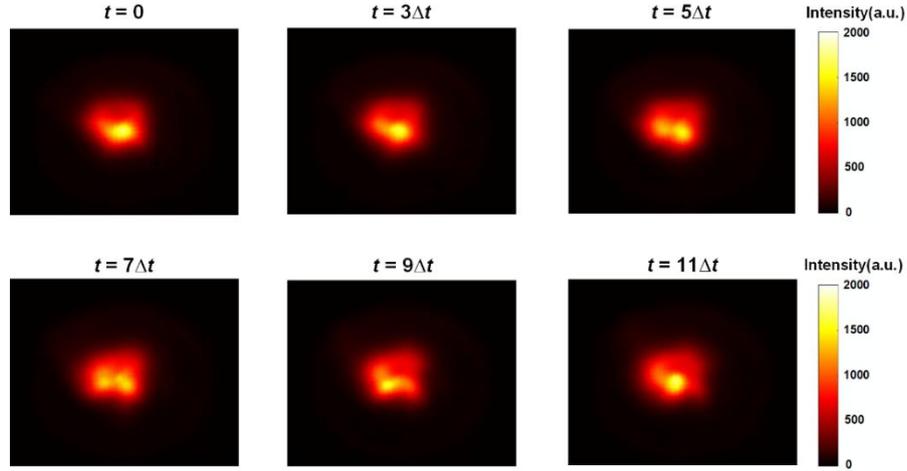

Fig. A2 Light emission intensity for the cross-section of the plasma column, captured by a fast camera. The exposure time is 7 μs, with a time interval of Δt=37 μs between consecutive snapshots. The plasma column (bright region) has a radius of approximately 5 cm.

*Appendix B: Rotation frequency of the helical structure* — In the non-neutral regime, the plasma can be treated as a cloud of net negative charges undergoing collective azimuthal rotation due to the $E_{eff}\times B$ drift, where an effective electric field, $E_{eff}$, primarily arises from the asymmetric charge distribution. A similar phenomenon has been reported in previous studies of the diocotron mode in pure electron plasmas [66-68]. Using the image charge method, the angular frequency of the helical structure can be estimated as

$$\omega \approx \frac{|\lambda|}{2\pi\varepsilon_0 BR^2}, \tag{B1}$$

where $\lambda = -\int e(n_e - n_i)dS$ is the linear net charge density, and $R$ is half of the transverse length. The derivation of Eq. (B1) can be found in Section VI of the Supplemental material [41].

For the case with p=1 mTorr, the parameters at y=10 cm were found to be $\lambda \approx -6.9\times 10^{-10}$ C/m and R=1.25 cm. Substituting these values into Eq. (B1) yields ω=8.1 MHz, which is qualitatively consistent with the simulated value of ω=5.6 MHz. This discrepancy is likely due to the shape of electron density not being ideally cylindrical, thereby introducing additional multipole effects.